\documentclass[notoc]{JHEP3}

\usepackage{epsfig}

 \def\dS{de Sitter} \def\adS{anti--\dS}
\def\RS{Randall--Sundrum} \def\dsp{\displaystyle}
\def\BEq{\begin{equation}}\def\EEq{\end{equation}} \def\DD{{\cal D}_2}
\def\BEa{\begin{eqnarray}}\def\EEa{\end{eqnarray}} \def\dd{{\rm d}}
\def\RR{{\cal R}} \def\pt{\partial} \def\LA{\Lambda} \def\DE{\Delta}
\def\textfrac#1#2{{\textstyle{#1\over#2}}}
\def\ph{\varphi} \def\rarr{\rightarrow}
\def\xh{x} \def\gh{\bar g} \def\Ll{\bar\lambda} \def\rh{\rho}
\def\rhf{1+{\rh^2\over a^2}}
\def\Rhf{\left(\dsp\rhf\right)}
 \def\de{\delta} \def\PH{\Phi} 
\def\FF{{\mathbf F}} \def\caseA{$\LA=0$, $\Ll=0$}
\def\caseB{$\LA=0$, $\Ll>0$}
 \def\e{{\rm e}}
\def\ff{{\cal F}} 
\def\Gfn{{\mathsf\Delta}} \def\Gkn{\Gfn_{k,n}} \def\rhd{(\rh,\rh')}
\def\Gknrh{\Gkn\rhd} \def\hdl{{\Gfn}_<} \def\hdg{{\Gfn}_>}

\def\goesas{\mathop{\sim}\limits} \def\WW{{\cal W}\!} \def\AA{{\cal A}}
\def\BB{{\cal B}} \def\Xsn{X,_{*n}}\def\Ysn{Y,_{*n}}
\def\scr{\scriptstyle}

\newcommand{\BbbR}{\mathbb{R}}
\newcommand{\BbbZ}{\mathbb{Z}}
\newcommand{\lowenspace}{{M_{\mathrm{low}}}}

\title{Brane worlds with bolts}

\author{Jorma Louko$^{\bf a}$
and
David L. Wiltshire$^{\bf b,c}$\\
$^{\bf a}$School of Mathematical Sciences,
University of Nottingham,
University Park,
Nottingham NG7 2RD,
United Kingdom\\
$^{\bf b}$Department of Physics and Mathematical Physics,
Adelaide University,
S.~A. 5005,
Australia\\
$^{\bf c}$\thanks{Present address}
Department of Physics and Astronomy,
University of Canterbury,
Private Bag 4800,
Christchurch,
New Zealand\\
E-mail: \email{jorma.louko@nottingham.ac.uk},
\email{dlw@phys.canterbury.ac.nz}}

\abstract{We construct a family of $(p+3)$--dimensional brane worlds in
which the brane has one compact extra dimension, the bulk has two
extra dimensions, and the bulk closes regularly at codimension two
submanifolds known as bolts. The $(p+1)$--dimensional low energy
spacetime $\lowenspace$ may be any Einstein space with an arbitrary
cosmological constant, the value of the bulk cosmological constant is
arbitrary, and the only fields are the metric and a bulk Maxwell
field.
The brane can be chosen to have positive tension, and the closure of
the bulk provides a singularity-free boundary condition for solutions
that contain black holes or gravitational waves in~$\lowenspace$. The
spacetimes admit a nonlinear gravitational wave whose
properties suggest that the Newtonian gravitational potential on a
flat $\lowenspace$ will behave essentially as the static potential of
a massless minimally coupled scalar field with Neumann boundary
conditions. When $\lowenspace$ is $(p+1)$--dimensional Minkowski with
$p\ge3$ and the bulk cosmological constant vanishes, this
static scalar potential is shown to have
the long distance behaviour characteristic
of~$p$ spatial dimensions. }


\preprint{$\hbox{hep-th/0109099, ADP-00-45/M93}\atop
\hbox{\jhep{0202}{2002}{007}}$}

\begin{document}

\section{Introduction}
\label{sec:intro}

The idea that our universe might be a 3--brane embedded in a
higher--dimensional spacetime with large bulk dimensions
\cite{Ak}--\cite{GW} has become the focus of intense research over the
past two years. A~pivotal observation was Randall and Sundrum's
discovery \cite{RS2} that a flat positive tension 3--brane can be
embedded in 5--dimensional \adS\ space in such a way that the static
gravitational potential on the brane reduces in the large distance
limit to the (3+1)-dimensional Newtonian potential with polynomial
corrections \cite{RS2}--\cite{ZK}. The subject has been
developed into several directions, including
black holes on branes
\cite{CHR}--\cite{emp-gregory-santos},
cosmological brane
worlds (see for example
\cite{bridgman-malik-wands} and the references therein),
and brane worlds with more than one extra dimensions
\cite{chodos-poppitz}--\cite{corradini-etal}.

In the \RS\ single-brane model
\cite{RS2} the unperturbed bulk spacetime is locally \adS.
One appealing consequence is the possibility to analyse
brane worlds in the context of $M$-theoretic AdS/CFT correspondence
\cite{GKR,ponton-poppitz,gubser-ads/cft,odintsov-cft,Haw-Her-Reall,BCR}.
What is less appealing is the presence of an \adS\ Killing horizon in
the bulk. This horizon tends to develop curvature
singularities upon addition of perturbations, as
has been found with black holes \cite{CHR,chamblin-reall-etal} and
gravitational waves~\cite{BCR,CG}. This raises questions about boundary
conditions that may need to be imposed in the bulk and about their
consequences on the brane. For example, in the higher-codimension brane
world scenarios of~\cite{gher-shapo,gher-roe-shapo},
the treatment of a bulk singularity affects the corrections to
Newton's law on the brane~\cite{ponton-poppitz}.

In this paper we introduce a family of brane world spacetimes in which the
\adS\ horizons of the \RS\ model
are replaced by bolts~\cite{GH}: totally geodesic codimension two
submanifolds at which a rotational Killing vector field vanishes.
The bulk closes regularly at the bolts,
and this closing provides a topological,
singularity-free boundary condition for
gravitational waves or any other perturbations one may wish to
consider.
The effective low energy spacetime $\lowenspace$ may be any Einstein
space, with any value of the low energy cosmological constant. The brane
is the product of $\lowenspace$ and one compact extra dimension, while
the bulk has two extra dimensions and solves the electrovacuum Einstein
equations with an arbitrary value of the bulk cosmological constant.
Solutions with positive brane tension exist for any values of the low
energy cosmological constant and the bulk cosmological constant, while
for certain values of these constants there also exist solutions with a
negative brane tension.

Related brane worlds with more than one extra dimension have been
presented in several contexts
\cite{chodos-poppitz}--\cite{corradini-etal}, and the general
constraints on these types of constructions are discussed
in~\cite{leblond-myers-winters}. Our main observation
is that a pure Einstein-Maxwell theory succeeds in regularly
closing the bulk for any values of the low energy cosmological constant
and the bulk cosmological constant.

As the extra dimensions in our model are compact, one expects that the
effective gravitational dynamics in $\lowenspace$ reduces to Einstein's
equations at length scales that are large compared with the characteristic
scales of the bulk.
In particular, when $\lowenspace$ is $(p+1)$--dimensional
Minkowski spacetime with $p\ge3$, one expects the Newtonian
gravitational potential in $\lowenspace$ to be proportional to
$-{|{\mathbf x} - {\mathbf x}'|}^{2-p}$, with corrections that vanish
exponentially when ${|{\mathbf x} - {\mathbf x}'|} \to\infty$. We shall
not attempt to verify this from a full linearized perturbation analysis,
but we shall show that when the bulk cosmological constant vanishes,
the corresponding result does hold for the static
potential of a massless
minimally coupled scalar field with Neumann boundary conditions on the
brane. We shall also present an exact nonlinear gravitational wave
solution on our brane world background and use it to argue that the
scalar field with Neumann boundary conditions is likely to capture
the essentials of the linearized gravitational field. This is known to
be the case for linearized perturbations of the \RS\ brane world
\cite{RS2}--\cite{ZK}.

We begin in section
\ref{sec:thickbrane} with a brief review of the
Einstein-Maxwell
thick brane world model
introduced by Gibbons and Wiltshire in 1987~\cite{GW}, on which our
construction is based. These thick branes are smooth codimension two
submanifolds, with a number of attractive features in their own right,
including localized chiral fermion modes on the brane, a mass gap for scalar
and tensor modes in the presence of a negative bulk cosmological constant,
and a mass gap for these modes
even without a bulk cosmological constant if the thick brane
has positive curvature. However, the major deficiency of the model is
the lack of a well-defined graviton zero mode that
would produce an effective low-dimensional Newton's law on the brane.

In section \ref{sec:surgery} we reinterpret the thick brane of Gibbons
and Wiltshire as a bolt at which the bulk spacetime closes, introduce
a thin brane between two such thick branes and interpret the thin brane
as the product of the low energy spacetime $\lowenspace$ with one compact
extra dimension. We show that the
field equations can be satisfied with a pure brane tension term on the
thin brane and analyse the constraints on the
the brane tension and the bulk magnetic field for given values of the
low energy cosmological constant and the bulk cosmological constant.
In particular, we show that when the low energy cosmological constant
vanishes, there is necessarily a nonvanishing bulk magnetic field.

In section \ref{sec:nonlinwave} we present an exact nonlinear gravitational
wave on the brane world background and show that the field
equations for this gravitational wave reduce to a massless minimally
coupled scalar field equation with Neumann boundary conditions on the
thin brane. In section \ref{sec:scalarfield} we analyse the static
potential of a massless minimally coupled scalar field when
$\lowenspace$ is $(p+1)$--dimensional
Minkowski spacetime with $p\ge3$ and the bulk cosmological constant
vanishes.
Section \ref{sec:discussion} is devoted to brief concluding remarks.

Our metric signature is $(-++\cdots)$. The spacetime
dimension is $p+3$ with $p\ge1$.

\section{The thick brane world}
\label{sec:thickbrane}

In this section we briefly review the thick brane
worlds of Gibbons and Wiltshire~\cite{GW}.
The idea is to make the low energy spacetime a
codimension two warped product submanifold of a higher--dimensional
curved spacetime with matter in the bulk,
such that the extra dimensions have
infinite volume\footnote{This model differs in a number of aspects from
previous models with extra dimensions of infinite volume studied in the
mid-1980s: the models of Akama \cite{Ak} and Rubakov and Shaposhnikov
\cite{RuSha} were by contrast based on the field theoretic trapping of matter
about a surface in a higher--dimensional {\it flat\/} spacetime. The model of
Visser \cite{Vis} was based on a 5--dimensional curved spacetime with a
diagonal metric whose time component was multiplied by a warp factor but
whose space components were not, resulting in a 4--dimensional
submanifold for the physical spacetime which did not possess an exact
Lorentz invariance.}.

The field equations are the $(p+3)$--dimensional Einstein-Maxwell
equations, derivable from the action
\BEq
S=\int\dd^{p+3}x\sqrt{-g}
\left\{ \frac1{2\kappa^2} \left(\RR-2\LA\right)
- \frac14 F_{ab}F^{ab}\right\},
\label{thickaction}
\EEq
where $F_{ab}$ is the field strength of the $U(1)$
gauge field.
The solutions are conveniently written as
\BEa
\dd s^2=r^2\gh_{\mu\nu}\dd\xh^\mu\dd\xh^\nu+{\dd r^2\over\DE}+
\DE\dd\ph^2,
\label{thickbulkG}
\\
\FF=
\frac{\left[p(p+1)\right]^{1/2} B}{\kappa r^{p+1}}
\,
\dd r\wedge\dd\ph
\,,
\label{thickF}
\EEa
where
\BEq
\DE
=
\Ll+{A\over r^p}-{B^2\over r^{2p}}
- \frac{2 \Lambda r^2}{(p+1)(p+2)}
\,,
\label{delta}
\EEq
$A$ and $B$ are real-valued constants
and $\gh_{\mu\nu}(\xh)$ is the metric on a
$(p+1)$--dimensional Einstein spacetime of signature $(-++\cdots)$,
\BEq
\bar R_{\mu\nu}= p\Ll\,\gh_{\mu\nu}\,.
\label{thickbraneG}
\EEq
We denote this $(p+1)$--dimensional Einstein spacetime
by~$\lowenspace$.
The numerical factor in (\ref{thickbraneG}) is chosen so that
$\Ll$ is equal to the Gaussian curvature of~$\lowenspace$.
Latin indices from the beginning of the alphabet
run over all bulk dimensions, $a,b=0,1,\dots,p+2$,
while Greek indices run over the dimensions of~$\lowenspace$,
$\mu,\nu=0,1,\dots,p$.

The bulk metric is thus the warped product of $\lowenspace$
and a 2--dimensional
space, $\DD$, and for $B\ne0$
there is a magnetic field in the bulk.
These solutions are dual by
double analytic continuation to solutions that generalize
electrically charged Reissner-Nordstr\"om spacetimes~\cite{GW}.

Among all the various cases -- which arise from a classification of the
zeros of $\DE$~\cite{GW} -- the spacetimes of interest as thick brane
worlds are those in which $\DD$ is geodesically complete and of infinite
total volume\footnote{The case in which $\DD$ was non--compact but of
finite volume and geodesically incomplete was studied previously by
Wetterich \cite{wetterich}.}. These are respectively: (i) \caseA;
(ii) \caseB; (iii) $\LA<0$, $\Ll$ arbitrary.
If the largest positive zero of $\DE(r)$ is at $r=r_0$ and
$\DE'(r_0)>0$,
where the prime denotes derivative with respect to~$r$,
then $\DE(r)$ is positive for $r_0 < r<\infty$.
Provided $\varphi$ has period $4\pi/\DE'(r_0)$,
the geometry can be regularly extended to the
$(p+1)$--dimensional totally geodesic
submanifold $r=r_0$, which consists of the fixed points of the Killing
vector $\pt/\pt\ph$ and is known as a ``bolt'' in the
terminology of~\cite{GH}. This submanifold is interpreted as the core
of the brane.

These thick brane world are generalizations of Melvin's magnetic
universe~\cite{Mel,Kip}. This is most readily seen in case~(i).
Taking $B\ne0$ and $A>0$ and defining
the coordinates $({\tilde x}^\mu, \rho, \phi)$ by
\BEa
{\tilde x}^\mu
&=&
\left(\frac{B^2}{A}\right)^{1/p}
x^\mu \, ,
\\
\rho
&=&
\frac{2}{pA}
\left(\frac{B^2}{A}\right)^{1/p}
r^{p}\sqrt{\DE} \, ,
\\
\phi
&=&
\frac{p A^2}{2 B^2}
\left(\frac{A}{B^2}\right)^{1/p}
\varphi \, ,
\EEa
the solution reads
\BEa \dd s^2&=&\Rhf^{2/p}
\left(\eta_{\mu\nu}\dd{\tilde x}^\mu\dd{\tilde x}^\nu
+\dd\rh^2\right)+{\rh^2\dd\phi^2\over\Rhf^2}\, ,
\label{flatbraneG}
\\
\FF&=&
\frac{2}{\kappa a}
\left(\frac{p+1}{p}\right)^{1/2}
\frac{\rh \, \dd\rh\wedge\dd\phi}{\Rhf^2}\, ,
\label{flatbraneF}
\EEa
where $\eta_{\mu\nu} = \hbox{diag}(-1,1,1,\cdots)$ and
\BEq
a=
\frac{2B}{pA}
\left(\frac{B^2}{A}\right)^{1/p} .
\EEq
The periodicity of $\varphi$ implies that $\phi$ has period $2\pi$,
and the solution is evidently regular at the bolt at $\rho=0$.
Melvin's magnetic universe is recovered for $p=1$.

The solutions are
stable \cite{GW} for much the same reasons as Melvin's solution is
stable~\cite{Kip}. Essentially one has an equilibrium configuration in
which the mutual repulsion of magnetic flux lines
is balanced by their gravitational attraction.

Generalizations including a bulk
dilaton were subsequently given by Gibbons and Maeda~\cite{GM}.
The properties of the non--compact space $\DD$ in these
generalizations are similar. Recent generalizations in the context of
$M$-theory are given in~\cite{saffin,GS,CHC,uranga}.

In \cite{GW} the problem of localising gravity and gauge fields on the
thick brane was considered in some detail. For case~(i),
it was shown that the bulk Dirac equation has zero mode solutions
that are confined to a neighbourhood of the brane and
move within it like massless $(p+1)$--dimensional chiral fermions.
Note that this localization is achieved by gravity and a bulk Abelian
gauge field alone. By contrast, the recent
mechanisms of \cite{fermions} for localising fermions on brane worlds
invoke matter fields in addition to gravity.

For a massless minimally coupled Klein-Gordon field, cases (ii) and
(iii) were found to have a mass gap, but in case (i) the mass spectrum
is continuous down to zero and does not give localization on the brane.
A perturbation analysis was not carried out for all the
gravitational modes, but a detailed analysis of nonlinear radial
gravitational perturbations in case (i) yielded a continuous mass
spectrum and supported the conclusion that the broad picture for the
gravitational modes is similar to that for the scalar modes.

The major deficiency of this thick brane model
is the absence of a
normalizable graviton zero mode that would
produce linearized $(p+1)$--dimensional Einstein gravity at the core of
the thick brane, and in particular produce the $(p+1)$--dimensional Newtonian
limit when $\lowenspace$ is flat. In the next section we will change the
viewpoint and use the solution (\ref{thickbulkG})--(\ref{thickbraneG})
to build a {\em thin\/} brane world in
which a normalizable graviton zero mode does exist.

\section{New brane world by brane surgery}
\label{sec:surgery}

We now add to the $(p+3)$--dimensional action (\ref{thickaction}) the term
\BEq
S_\mathrm{brane} = - \lambda \int_\mathrm{brane}
\dd^{p+2}x\sqrt{-{\mathrm{det}\left(g_{{\hat\mu}{\hat\nu}}\right)}}
\,,
\label{braneaction}
\EEq
where the (new)
brane is a timelike hypersurface of codimension one,
$g_{{\hat\mu}{\hat\nu}}$ is the induced metric on this brane,
and the brane tension $\lambda$ is a
nonvanishing constant.
The metric and $F_{ab}$ are assumed continuous across the brane but
their first derivatives may be discontinuous.
Einstein's equations at the
brane amount to the Israel junction conditions
\cite{israel-shell,MTW,barrabes-israel,mukohyama-pertgen}
\BEq
K_{{\hat\mu}{\hat\nu}}^+ - K_{{\hat\mu}{\hat\nu}}^-
= - \frac{\kappa^2 \lambda}{p+1} g_{{\hat\mu}{\hat\nu}}
\,,
\label{israel-junction}
\EEq
where
$K_{{\hat\mu}{\hat\nu}}^{\pm}$
are the extrinsic curvatures of the brane in the geometries on the
two sides, with respect to the normal that points from the
``$-$'' side to the ``$+$'' side.\footnote{We follow the
convention of \cite{barrabes-israel,mukohyama-pertgen} for the
sign of the extrinsic curvature.} The hatted Greek indices run over
the $p+2$ dimensions on the brane.

To construct the solution with a thin brane,
we begin with the bulk solution
(\ref{thickbulkG})--(\ref{thickbraneG}).
Let $r=r_0$ be a positive zero of~$\DE$, such that $\DE'(r_0)\ne0$.
We
make
$\varphi$ periodic with period $4\pi/|\DE'(r_0)|$.
If $\DE'(r_0)>0$, we choose a constant $r_* > r_0$ such that
$\DE>0$ for $r_0 < r \le r_*$.
If $\DE'(r_0)<0$,
we similarly choose a constant, also denoted by~$r_*$,
so that $0 < r_* < r_0$ and
$\DE>0$ for $r_* \le r < r_0$.
We denote
the spacetime in which
$r$ ranges from $r_0$ to $r_*$ by~${\cal M}_-$.
$r=r_0$ is a bolt in the interior of ${\cal M}_-$,
and the geometry is regular there,
while $r=r_*$ is a boundary of ${\cal M}_-$, consisting of $\lowenspace$
(with metric multiplied by~$r_*^2$)
and a spacelike circle with
circumference $4\pi \DE^{1/2}(r_*)/|\DE'(r_0)|$.

Let ${\cal M}_+$ be a copy of ${\cal M}_-$.
We glue ${\cal M}_-$ and ${\cal M}_+$ together
at $r=r_*$, using the identity
gluing in $\lowenspace$ and
choosing the gluing in $\varphi$ so that
$F_{ab}$ is continuous.
An embedding diagram of the resulting spacetime~${\cal M}$, with
$\lowenspace$ suppressed,
is shown in
Figure~\ref{fig:embedding}.

\FIGURE{\epsfig{file=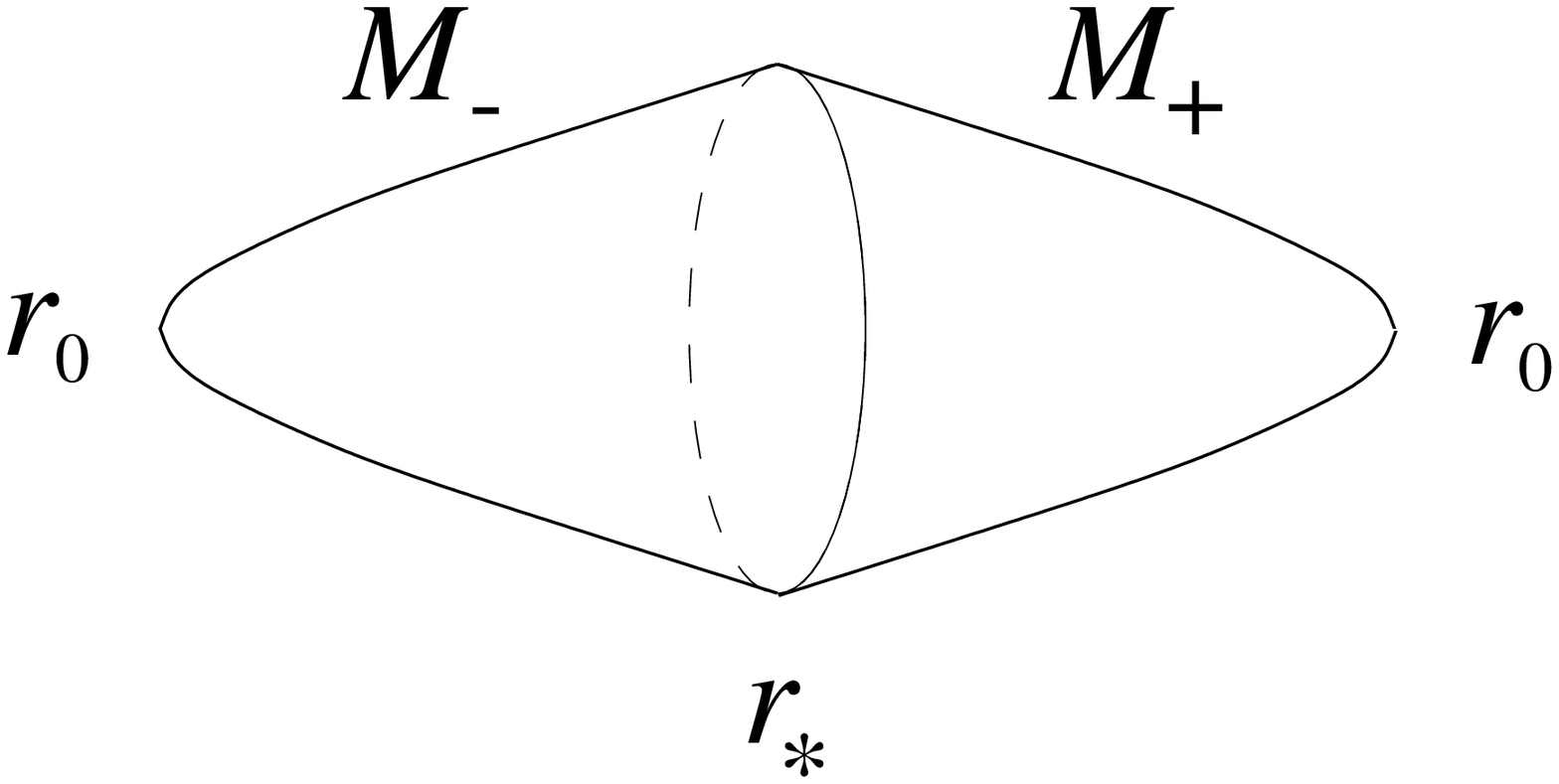,width=9cm}
\caption{An embedding of the $r$ and $\varphi$ dimensions of
${\cal M}$ into~$\BbbR^3$.}%
\label{fig:embedding}}

We wish to make ${\cal M}$ into a solution with the thin
brane action~(\ref{braneaction}), such that the thin brane is at the
junction $r=r_*$. It is easily verified that Maxwell's equations are
satisfied at the junction. As the extrinsic curvature of a
constant $r$ hypersurface in the metric
(\ref{thickbulkG}) is $\frac12\DE^{1/2}
\partial_r g_{{\hat\mu}{\hat\nu}}$, with respect to the normal
pointing towards increasing~$r$, the gravitational
junction conditions
(\ref{israel-junction}) amount to the pair
\BEa
\sqrt{\DE(r_*)}
&=& \frac{\epsilon \kappa^2 \lambda r_*}{2(p+1)}
\,,
\label{branejunction1}
\\
\DE'(r_*)
&=& \frac{\epsilon \kappa^2 \lambda \sqrt{\DE(r_*)} }{p+1}
\,,
\label{branejunction2}
\EEa
where $\epsilon = \mathrm{sgn}(r_* - r_0)$.
Eliminating $\lambda$ yields
\BEq
\frac{\partial}{\partial r}
\left. \left(\frac{\DE}{r^2}\right)\right|_{r = r_*} =0
\,.
\label{rstar-eq}
\EEq
The solutions are thus obtained by solving~(\ref{rstar-eq}),
subject to the conditions on $r_0$ and $r_*$ stated above,
and then determining $\lambda$ from (\ref{branejunction1})
or~(\ref{branejunction2}). Note
that $\lambda$ has the same sign as $r_* - r_0$.

An elementary analysis shows that solutions with positive brane tension
exist for any $\Ll$ and~$\Lambda$.
The restrictions on $B$ and the sign of $\lambda$ for given
$\Ll$ and $\Lambda$ are collected in Table~\ref{tab:zoo}.
Note that solutions with $F_{ab}=0$ exist only when $\Ll>0$.

\TABLE{
\begin{tabular}{c|c@{\hskip 1 cm}c@{\hskip 1 cm}c}
& $\Lambda<0$ & $\Lambda=0$ & $\Lambda>0$ \\
\hline
$\Ll<0$ & $B\ne0$ & $B\ne0$ & $B\ne0$ \\
$\Ll=0$ & $\lambda>0$, $B\ne0$ &$\lambda>0$, $B\ne0$ & $B\ne0$ \\
$\Ll>0$ & $\lambda>0$ & $\lambda>0$ & none \\
\end{tabular}
\caption{Restrictions on $B$ and the sign of $\lambda$
for given $\Ll$ and~$\Lambda$.}
\label{tab:zoo}}

We note in passing that if ${\cal M}_-$ and ${\cal M}_+$ are glued
together so that $F_{ab}$ changes sign at the junction, we obtain as
above
solutions to the Einstein-Maxwell equations with an external
current proportional
to $\partial/\partial\varphi$ on the brane. A~variational principle
producing this external current
is obtained by including the ``cosmological current'' term
\BEq
\int \dd^{p+3}x \, \tilde{J^a} A_a
\,,
\label{currentcouplingterm}
\EEq
where $\tilde{J^a}$ is a
prescribed
conserved current density with support only on the brane. However, these
solutions have the unphysical feature that the brane's stress-energy
tensor, which is unaffected by the term~(\ref{currentcouplingterm}),
contains no contribution from charge carriers on the
brane.\footnote{We thank Eric Poisson for correspondence on this
point.}

We seek to interpret ${\cal M}$ as a brane world in which the spacetime
of low energy physics is the $(p+1)$--dimensional Einstein spacetime
$\lowenspace$ on the thin brane. The brane thus has one extra dimension,
which is compact, and the two extra dimensions in the bulk close regularly
at two thick branes of codimension two.
For this interpretation to be viable, the Newtonian potential between
sources concentrated on the thin brane and constant in $\varphi$
needs to reduce to the $(p+1)$--dimensional Newton's law in
the Newtonian limit. The next two sections will present evidence that
this is likely to be the case.

\section{Non--linear gravitational waves on the brane world}
\label{sec:nonlinwave}

Chamblin and Gibbons \cite{CG} have constructed plane polarized
nonlinear gravitational waves on the \RS\ spacetime and its thick brane
generalizations. The construction utilizes the solution-generating
technique of Garfinkle and Vachaspati~\cite{GV}, which works for any
electrovacuum spacetime that admits a hypersurface-orthogonal null
Killing vector. We now show how this technique can be applied to the
spacetimes of sections \ref{sec:thickbrane} and~\ref{sec:surgery},
assuming that $\lowenspace$ admits a hypersurface-orthogonal null
Killing vector.

Let $k^\mu$ be a
hypersurface-orthogonal null Killing vector
on~$\lowenspace$. We raise and lower the index on $k^\mu$ with
$\gh_{\mu\nu}$. As discussed in~\cite{GV}, there then
exists on $\lowenspace$
at least locally a scalar ${\bar f}$ such that
$\partial_{[\mu} k_{\nu]} =
k_{[\nu} \partial_{\mu]} {\bar f}$
and
$k^\mu \partial_\mu {\bar f}=0$.

We extend $k^\mu$ into a vector field $l^a$ on
the spacetime (\ref{thickbulkG})--(\ref{thickbraneG})
by the
natural extension, $l^a = (k^\mu, 0, 0)$.
We raise and lower the index on $l^a$ with the full spacetime
metric~(\ref{thickbulkG}).
It is straightforward to verify
that $l^a$ is a
nonvanishing hypersurface-orthogonal null Killing vector.
Further, $l^a$ satisfies
$\partial_{[a} l_{b]} =
l_{[b} \partial_{a]} \left( f + 2\ln r \right)$
and
$\ell^a \partial_a \left( f + 2\ln r \right) =0$, where $f$ is the
pull-back of ${\bar f}$ to~(\ref{thickbulkG}).

As $l^a F_{ab} =0$ and $\pounds_l F_{ab}=0$, $l^a$ satisfies the
conditions of the solution-generating technique of~\cite{GV}. Thus let us
take $H$ to be a scalar function on
the spacetime (\ref{thickbulkG})--(\ref{thickbraneG}),
satisfying $l^a \partial_a
H=0$ and $\nabla_a \nabla^a H =0$, where $\nabla_a$ is the covariant
derivative in the metric~(\ref{thickbulkG}).
It follows \cite{GV} that
adding to (\ref{thickbulkG}) the term
\BEq
r^2 H \e^{-f} \, k_\mu k_\nu
\dd\xh^\mu\dd\xh^\nu
\label{waveterm}
\EEq
gives a solution to the Einstein-Maxwell
equations. This solution can be interpreted as a nonlinear gravitational
wave on the spacetime
(\ref{thickbulkG})--(\ref{thickbraneG}),
travelling in the direction of~$l^a$.

To include the wave term (\ref{waveterm})
in the thick brane worlds
of section~\ref{sec:thickbrane}, one needs to check that the geometry
remains regular at the bolt. This is equivalent to the requirement of
regularity of $H$ as a scalar field
on the spacetime without the term~(\ref{waveterm}).
To include the wave term
in the thin brane worlds of section~\ref{sec:surgery}, one needs in
addition to check that the junction conditions
(\ref{israel-junction}) are satisfied at the thin brane.
As the only
$r$--dependence in (\ref{waveterm}) is in the factor~$r^2 H$,
the junction conditions consist of~(\ref{branejunction1}),
(\ref{branejunction2}) and
\BEq
\sqrt{\DE(r_*)}
\,
\partial_r \! \left. (r^2 H) \vphantom{A^A_A}\right|_{r=r_*} =
\frac{\epsilon \kappa^2 \lambda r_*^2 H(r_*)}{p+1}
\, .
\label{branejunction3}
\EEq
Using (\ref{branejunction1}), (\ref{branejunction3}) reduces to
$\partial_r H \bigr|_{r=r_*} =0$. When viewed as a scalar field on the
thin brane spacetime without the term~(\ref{waveterm}), $H$
thus obeys the Neumann boundary condition at the thin brane.

As an example, set
$\gh_{\mu\nu} = \eta_{\mu\nu}$. Adopting double null
coordinates $(u,v, x_\perp^k)$, where $k = 2,\ldots,p$,
and choosing $k^\mu =
(\partial_v)^\mu$, the solution reads
\BEq
\dd s^2=r^2\left[-\dd u\dd v +H(u, x_\perp^k,r,\ph)\dd u^2
+\delta_{ij} \dd x_\perp^i \dd x_\perp^j \right]
+{\dd r^2\over\DE}+\DE\dd\ph^2
\,,
\label{wave-on-eta}
\EEq
where $\DE(r)$ is given by
(\ref{delta}) with $\Ll=0$.
The construction is thus global on~$\lowenspace$.
The scalar wave equation for $H$ reads explicitly
\BEq
H,_{rr} + \left({p+1\over r}+{\DE,_r\over\DE}\right)H,_{r}+{H,_{\ph\ph}
\over\DE^2}+{\delta^{ij}H,_{ij} \over r^2\DE}=0.
\label{Hcondition}
\EEq
Note that $H$ does not depend on $v$ but its dependence on $u$ is
arbitrary. The linearized limit of the solution can be discussed as
in~\cite{CG}. In particular,
$H = h_{ij}(u) x_\perp^i x_\perp^j$ is clearly a solution, it satisfies
$\delta^{ij}H,_{ij}=0$, and its linearized limit is
analogous to the famous normalizable massless
mode in the \RS\ spacetime~\cite{RS2}.
This is the normalizable graviton
zero mode we promised at the end of section~\ref{sec:thickbrane}.

We note that in the context of the thick brane worlds of Gibbons and
Wiltshire,
the scalar field spectral analysis of \cite{GW} applies directly
to
the nonlinear gravitational wave~(\ref{wave-on-eta}).
To see this explicitly, we decompose $H$ as
\BEq
H=\ff(r)\e^{i{\tilde n}\ph}e^{ik_j x_\perp^j}
\,,
\label{HFourier}
\EEq
where $k_j$ may depend on~$u$. As $\ph$ has period $4\pi/\DE'(r_0)$,
regularity at the bolt requires ${\tilde n} = \textfrac12\DE'(r_0) n$,
$n \in \BbbZ$. For modes that do not exhibit faster-than light
propagation on the thick brane, we need $\delta^{ij}H,_{ij} = m^2 H$,
where $m^2 \ge0$: this is achieved if $k_j$ is purely imaginary.
The radial function $\ff(r)$ then obeys
\BEq
-{\dd\hphantom{r}\over\dd r}
\left[r^{p+1}\DE{\dd\ff\over\dd r}\right]
-m^2r^{p-1}\ff+{\tilde n}^2r^{p+1}{\ff\over\DE}=0
\,,
\label{radialmode}
\EEq
which is identical to Eq.\ (6.1c) in~\cite{GW}.
The results of \cite{GW} imply that if $\Lambda<0$,
the requirement that $H$ vanishes at $r\to\infty$
makes the spectrum of $m^2$
discrete and positive. If $\Lambda=0$, on the other hand, the
spectrum of $m^2$ for ${\tilde n}=0$ is the positive real
line.

As a second example, we take
$\gh_{\mu\nu}$ to be the metric of \adS\ space.
Adopting double null horospherical
coordinates $(z, u,v, x_\perp^k)$, where $z>0$ and $k = 3,\ldots,p$,
and
choosing $k^\mu =
(\partial_v)^\mu$, the solution reads
\BEq
\dd s^2= \frac{r^2}{z^2}\left[\dd z^2 -\dd u\dd v
+H(u, x_\perp^k,r,\ph)\dd u^2
+\delta_{ij} \dd x_\perp^i \dd x_\perp^j \right]
+{\dd r^2\over\DE}+\DE\dd\ph^2
\,,
\label{wave-on-adS}
\EEq where $\DE(r)$ is given by (\ref{delta}) with $\Ll=-1$. However, the
horospherical coordinates are not global on \adS\ space (see for
example~\cite{BKL}), and generic solutions for $H$ develop singularities
on the Killing horizons at $z\to0$~\cite{BCR,CG,podolsky}\footnote{However,
the Killing vector $\partial_v$
may be continued into an everywhere nonvanishing null Killing vector
on \adS\ space. In the embedding of \adS\ space as the hyperboloid
$-\alpha^2 = \eta_{AB} X^A X^B$ in flat space with the metric $\dd s^2
= \eta_{AB} \dd X^A \dd X^B$, where $\eta_{AB} =
\hbox{diag}(-1,-1,1,\cdots)$, $\partial_v$ can be written as
$(m_{A}n^{B} - n_{A}m^{B})X^A \partial_B$, where $m^A$ and $n^A$ are
two constant orthogonal null vectors. We thank John Barrett for this
observation.}.

\section{Massless scalar field on the brane world}
\label{sec:scalarfield}

In this section we analyse the static potential for a massless minimally
coupled scalar field on the brane world of section \ref{sec:surgery}
with $\gh_{\mu\nu} = \eta_{\mu\nu}$ and $\Lambda=0$.
As $\Ll=0$, Table \ref{tab:zoo} shows that the magnetic field is
necessarily nonzero.
We work in the coordinates in which
${\cal M}_\pm$ are given by
(\ref{flatbraneG}) and~(\ref{flatbraneF}).
The bolt on each side is at $\rho=0$,
and the brane is at $\rho=\rho_*$, where
\BEq
\rh_*=\left(p\over p+2\right)^{1/2}|a|
\,.
\label{rhostar}
\EEq
From now on we drop the tildes from the coordinates ${\tilde
x}^\mu$ in~(\ref{flatbraneG}).

For simplicity of presentation, we consider a scalar field on ${\cal
M}_-$ with Neumann boundary conditions at $\rho=\rho_*$, and we evaluate
the static potential of this field
between two points at $\rho=\rho_*$. The static
potential on the thin brane of section \ref{sec:surgery} is obtained by
multiplying this result by two. Our technique closely follows that
of~\cite{GKR}.

\subsection{The propagator}
\label{subsec:propagator}

The action of a massless minimally coupled scalar field $\PH$ on
${\cal M}_-$ reads
\BEq
S_\PH
=
-\textfrac12
\int\dd^{p+3}x \, (-g)^{1/2}
\, (\nabla_a\PH) (\nabla^a\PH)
\,.
\EEq
The Green's function, $\Gfn_{p+3}$, obeys
\BEa
\nabla_a \nabla^a \Gfn_{p+3}
&=&
{\de^{p+1}(x-x')\de(\rh-\rh')\de(\phi-\phi')\over
\sqrt{-g}}\nonumber\\ &=& {\de^{p+1}(x-x')\de(\rh-\rh')\de(\phi-\phi')\over
\rh\Rhf^{2/p}}
\,.
\label{Greendef}
\EEa
The Neumann boundary condition at $\rh = \rh_*$ is
\BEq
\left.\pt_\rh\Gfn_{p+3}\right|_{\rh=\rh_*}=0.
\label{Neugen}
\EEq

In order to solve~(\ref{Greendef}), we Fourier decompose $\Gfn_{p+3}$ as
\BEq
\Gfn_{p+3}(x,\rh,\phi;x',\rh',\phi')
= \int{\dd^{p+1}k\over{(2\pi)}^{p+2}}
\e^{ik_\mu(x^\mu-{x'}\vphantom{\scr x}^\mu)}\sum_{n=-\infty}^{\infty}
\e^{in(\phi-\phi')}\Gknrh
\EEq
and substitute in (\ref{Greendef}) to obtain
\BEq
\left[{1\over\rh}\pt_\rh\left(\rh\pt_\rh\right)+q^2-{n^2\over\rh^2}
\Rhf^{2(1+p)/p}\right]\Gknrh={\de(\rh-\rh')\over\rh}
\,,
\label{SLn}
\EEq
where $q^2=-k_\mu k^\mu$. The index on $x^\mu$ and $k^\mu$ is raised
and lowered with $\eta_{\mu\nu}$.
The boundary condition (\ref{Neugen}) reads
\BEq
\left.\pt_\rh\Gknrh\right|_{\rh=\rh_*}=0.
\label{Neu}
\EEq

For $\rh\ne\rh'$, Eq.\ (\ref{SLn}) is recognized as a Sturm-Liouville
equation that reduces to Bessel's equation as $\rh\rarr0$. For each~$n$,
we choose for the homogeneous equation
linearly independent solutions $X_n(\rh)$ and $Y_n(\rh)$
that at $\rh\to0$ have the asymptotic behaviour
\BEa
X_n(\rh)&\goesas&J_n(q\rh)
\,,
\label{Xn-asymptotics}
\\
Y_n(\rh)&\goesas&N_n(q\rh)
\,,
\EEa
where $J_n$ and $N_n$ are respectively Bessel functions of the
first and second kind.
The Wronskian of $X_n(\rh)$ and $Y_n(\rh)$ then
satisfies
\BEq
\WW\left[X_n(\rh),Y_n(\rh)\right]\equiv X_n\pt_\rh Y_n-Y_n\pt_\rh X_n=
{2\over\pi\rh}\,.
\label{Wron}
\EEq
As the homogeneous equation for $n=0$ is exactly Bessel's equation,
we have
\BEa
X_0(\rh)&=&J_0(q\rh)
\,,
\\
Y_0(\rh)&=&N_0(q\rh)
\,.
\EEa

We denote the solution to (\ref{SLn}) by $\hdl\rhd$ for $\rh<\rh'$ and
$\hdg\rhd$ for $\rh>\rh'$.
$\hdl$~and $\hdg$ are each linear combinations of
$X_n(\rh) $ and~$Y_n(\rh)$, determined by the matching conditions
arising from (\ref{SLn}) and the boundary conditions at $\rh=0$ and
$\rh=\rh_*$. The boundary condition at $\rh=0$ is regularity of the
solution as a function on the spacetime at the bolt,
which
excludes~$Y_n(\rh)$, and the boundary condition at $\rh=\rh_*$
is~(\ref{Neu}).
These boundary conditions imply
\BEa
\hdl\rhd
&=&
\AA(\rh')X_n(\rh)
\,,
\label{sol0}
\\
\hdg\rhd
&=&
\BB(\rh')\left[\Ysn X_n(\rh)-\Xsn Y_n(\rh)\right]
\,,
\label{sol1}
\EEa
where
\BEq \Xsn\left.\equiv{\dd X_n\over\dd\rh}\right|_{\rh=\rh_*},
\qquad \Ysn\left.\equiv{\dd Y_n\over\dd\rh}\right|_{\rh=\rh_*}.
\label{sol2}
\EEq
The matching conditions from (\ref{SLn}) read
\BEa
\left.\hdl\right|_{\rh=\rh'}&=&\left.\hdg\right|_{\rh=\rh'}
\,,
\label{match1}
\\
\pt_\rh\left.\left(\hdg-\hdl\right)\right|_{\rh=\rh'}&=&{1\over\rh'}
\,,
\label{match2}
\EEa
and when applied to (\ref{sol0}) and (\ref{sol1}) they imply
\BEa
&&
\AA(\rh')X_n(\rh')
=
\BB(\rh')
\left[\Ysn X_n(\rh')-\Xsn Y_n(\rh')\right]
\,,
\label{match3}
\\
&&
\BB(\rh')
\left[\Ysn \partial_{\rh'}X_n(\rh')-\Xsn \partial_{\rh'}Y_n(\rh')\right]
- \AA(\rh') \partial_{\rh'}X_n(\rh')
={1\over\rh'}
\,.
\label{match4}
\EEa
With the help of (\ref{Wron}), equations (\ref{match3}) and
(\ref{match4})
can be solved for $\AA(\rh')$ and $\BB(\rh')$ provided $\Xsn\ne0$, with
the result
\BEq
\Gkn={-\pi X_n(\rh_<)\over2\Xsn}\left[\Ysn X_n(\rh_>)
-\Xsn Y_n(\rh_>)\right],
\label{sol}
\EEq
where $\rh_>$ ($\rh_<$) denotes the greater (lesser) of $\rh$
and~$\rh'$.
We thus have
\BEa
\Gfn_{p+3}(x,\rh,\phi;x',\rh',\phi')
&=&
\dsp\int{\dd^{p+1}k\over4{(2\pi)}^{p+1}}
\e^{ik_\mu(x^\mu-{x'}\vphantom{\scr x}^\mu)}
\times
\nonumber
\\
&&
\ \ \
\times
\sum_{n=-\infty}^{n=\infty} \e^{in(\phi-\phi')}X_n(\rh_<)
\Biggl[Y_n(\rh_>)-{\Ysn\over\Xsn}X_n(\rh_>)\Biggr]
\,.
\nonumber
\\
\label{fullpropagator}
\EEa

The choice of the Green's function can be specified in
(\ref{fullpropagator}) by a prescription at the
zeros of~$\Xsn$, where the poles of the integrand are. The
retarded Green's function is obtained by the prescription $q^2
\to (k^0 + i\epsilon)^2 - {\mathbf{k}}^2$, $\epsilon \to 0_+$.

\subsection{Static potential on the brane}
\label{subsec:staticpot}

We wish to recover from (\ref{fullpropagator}) the static potential at
$\rh=\rh'=\rh_*$. We assume from now on that $p\ge3$.

When $\rh=\rh'=\rh_*$, (\ref{fullpropagator}) reduces with the help of
(\ref{Wron}) to
\BEa
\Gfn_{p+3}(x,\rh_*,\phi;x',\rh_*,\phi')=\dsp{1\over\rh_*}\int{\dd^{p+1}k
\over{(2\pi)}^{p+2}}
\,
\e^{ik_\mu(x^\mu-{x'}\vphantom{\scr x}^\mu)}
&&
\!\!
\Biggl\{{2\over q^2\rh_*}-{J_2(q\rh_*)
\over qJ_1(q\rh_*)}\nonumber\\
&&
\ \ \
-\sum_{n\ne0} \e^{in(\phi-\phi')}{X_n(\rh_*)
\over\Xsn}\Biggr\}.\nonumber\\
\label{smallpropagator}
\EEa
We have
isolated in (\ref{smallpropagator})
the $n=0$ term from the rest and used Bessel
function identities to write this term in a way that explicitly shows
its pole structure near $q=0$.

The static potential is obtained by integrating the retarded Green's
function over the time difference, $t-t'$. As the retarded Green's
function is nonzero only for $t-t'>0$, it is convenient to do
this in
(\ref{smallpropagator}) under the $\dd^{p+1}k$ integral, with the result
\BEa
V({\mathbf x}, \phi; {\mathbf x}', \phi')
=
\dsp{1\over\rh_*}
\int
\frac{\dd^p {\mathbf{k}}}{{(2\pi)}^{p+2}}
\int_{-\infty}^{\infty}
\frac{\dd k^0}{i (k^0 - i\epsilon)}
\,
\e^{i \mathbf{k} \cdot ({\mathbf x} - {\mathbf x}')}
&&
\!\!
\Biggl\{{2\over q^2\rh_*}-{J_2(q\rh_*)
\over qJ_1(q\rh_*)}\nonumber\\
&&
\ \ \
-\sum_{n\ne0}^{\infty}\e^{in(\phi-\phi')}{X_n(\rh_*)
\over\Xsn}\Biggr\}
\,,
\nonumber
\\
\label{staticpotential}
\EEa
where $q^2
= (k^0 + i\epsilon)^2 - {\mathbf{k}}^2$ and $\epsilon \to 0_+$.

In (\ref{staticpotential}) consider the term in the integrand
proportional to $(k^0 - i\epsilon)^{-1}q^{-2}$. We close
the $k^0$ contour in the upper half-plane and do the integral
by residues. As the only pole within the contour is
at $k^0 = i\epsilon$,
the contribution to $V({\mathbf x}, \phi; {\mathbf x}', \phi')$ is
\BEa
V_{0,\mathrm{Mink}}({\mathbf x}; {\mathbf x}')
&=&
- \frac{1}{\pi\rh_*^2}
\int
\frac{\dd^p {\mathbf{k}}}{{(2\pi)}^p}
\frac{\e^{i \mathbf{k} \cdot ({\mathbf x} - {\mathbf
x}')}}{{\mathbf{k}}^2}
\nonumber
\\
&=&
- \frac{\Gamma(\frac{p}{2} - 1)}{4 \pi^{(p/2)+1} {\rh_*^2}
\, {|{\mathbf x} - {\mathbf x}'|}^{p-2}}
\nonumber
\\
&=&
- \frac{1}{\pi (p-2) \Omega_{p-1} {\rh_*^2}
\, {|{\mathbf x} - {\mathbf x}'|}^{p-2}}
\,,
\label{vnoughtmink}
\EEa
where $\Omega_{p-1}$ is the volume of the $(p-1)$--sphere. As is
clear already from (\ref{smallpropagator}),
$V_{0,\mathrm{Mink}}({\mathbf x}; {\mathbf x}')$
is proportional to the static potential of a free massless field
in $(p+1)$--dimensional
Minkowski space. Note from (\ref{flatbraneG})
and (\ref{rhostar}) that the spatial proper distance at $\rh = \rh_*$ is
${[2(p+1)/(p+2)]}^{1/p} {|{\mathbf x} - {\mathbf x}'|}$.

Let us then consider the remaining part of the
$n=0$ term in~(\ref{staticpotential}). We now close the
$k^0$ contour in the lower half-plane. As all the zeros of $J_1$ are
real, the poles within the
contour are at $q = \pm j_{1,s}/\rh_*$,
where $j_{1,s}$, $s = 1,2,\ldots$ are the positive zeros of~$J_1$. The
contribution to $V({\mathbf x}, \phi; {\mathbf x}', \phi')$ is therefore
\BEa
V_{0,\mathrm{Y}}({\mathbf x}; {\mathbf x}')
&=&
\frac{1}{\pi\rh_*^2}
\int
\frac{\dd^p {\mathbf{k}}}{{(2\pi)}^p}
\,
\e^{i \mathbf{k} \cdot ({\mathbf x} - {\mathbf x}')}
\sum_{s=1}^\infty
\frac{J_2(j_{1,s})}{J_1'(j_{1,s})} \times
\frac{1}{\mathbf{k}^2 + {(j_{1,s}/\rh_*)}^2}
\nonumber
\\
&=&
\frac{1}{\pi {(2\pi)}^{p/2} \rh_*^2}
\,
\sum_{s=1}^\infty
\frac{J_2(j_{1,s})}{J_1'(j_{1,s})}
{\left(\frac{j_{1,s}}{ \rh_* {|{\mathbf x} - {\mathbf x}'|}}
\right)}^{(p/2) -1}
K_{(p/2) -1} \left(j_{1,s}
{|{\mathbf x} - {\mathbf x}'|} / \rh_* \right)
\,,
\nonumber
\\
\label{vzerokk}
\EEa
where $K$ is the modified Bessel function of the second kind.
At large ${|{\mathbf x} - {\mathbf x}'|}$ the terms in
$V_{0,\mathrm{Y}}({\mathbf x}; {\mathbf x}')$
decay as
${|{\mathbf x} - {\mathbf x}'|}^{(1-p)/2}
\exp\left(-j_{1,s} {|{\mathbf x} - {\mathbf x}'|} /
\rh_*\right)$,
and $V_{0,\mathrm{Y}}({\mathbf x}; {\mathbf x}')$ is
thus of Yukawa type.
The dominant term is $s=1$, for which $j_{1,1}
\approx 3.83$~\cite{abra-stegun}.

For the $n\ne0$ terms in~(\ref{staticpotential}), we do not have an
expression for $X_n$ in terms of known functions. However, the
differential equation (\ref{SLn}) and the boundary conditions at $\rh=0$
and $\rh=\rh_*$ imply that the zeros of $\Xsn$ occur precisely when
$q^2$ is an eigenvalue of a self-adjoint differential
operator~\cite{dunfordII,reed-simonII}. Bringing the operator to a
standard form shows that the spectrum is purely discrete (Ref.\
\cite{dunfordII}, Theorem XIII.7.17), and a simple estimate using the
differential equation (\ref{SLn}) and the asymptotic expression
(\ref{Xn-asymptotics}) shows that there are no eigenvalues with
$q^2\le0$. This means that the only zeros of $\Xsn$ in the complex $q$
plane are at discrete positive values of~$q^2$. If the
asymptotic behaviour of $X_n(\rh_*)/\Xsn$ at large complex $q$ is such
that closing the $k^0$ contour in (\ref{staticpotential}) in the lower
half-plane can be justified, it follows as with (\ref{vzerokk}) that the
contribution to $V({\mathbf x}, \phi; {\mathbf x}', \phi')$ is
exponentially suppressed as ${|{\mathbf x} - {\mathbf x}'|}\to\infty$.

In conclusion, the static potential $V({\mathbf x}, \phi; {\mathbf x}',
\phi')$ consists of the effective $(p+1)$--dimensional term
$V_{0,\mathrm{Mink}}({\mathbf x}; {\mathbf x}')$ (\ref{vnoughtmink})
plus corrections that, subject to our
technical assumptions about the
$n\ne0$ modes, are exponentially suppressed as
${|{\mathbf x} - {\mathbf x}'|}\to\infty$.

\section{Discussion}
\label{sec:discussion}

We have presented a family of $(p+3)$--dimensional brane worlds in which
the brane has one compact extra dimension, the bulk has two extra
dimensions, and the bulk is compact, closing regularly at bolts where
a rotational Killing vector vanishes. The
spacetimes solve the $(p+3)$--dimensional Einstein-Maxwell equations,
with an arbitrary bulk cosmological constant, and the field
equations at the brane come from a brane tension that can be chosen
positive. The low energy spacetime $\lowenspace$ may be any
$(p+1)$--dimensional Einstein space. When $\lowenspace$ is $(p+1)$
Minkowski with $p\ge3$ and
the bulk cosmological constant vanishes, we showed that
a massless minimally coupled scalar field satisfying the
Neumann boundary conditions on the brane has
a static potential on the brane with the
long distance behaviour $-{|{\mathbf x} - {\mathbf x}'|}^{2-p}$,
characteristic of $p$ spatial dimensions. We
did not attempt a direct calculation of the Newtonian gravitational
potential on the brane, but we presented exact nonlinear gravitational
wave solutions whose field equations reduce to those of
a massless scalar field with Neumann boundary conditions at the brane.
Our scalar field results therefore suggest that when
$\lowenspace$ is Minkowski, the long distance
behaviour of the Newtonian
potential on the brane should also be
characteristic of $p$ spatial
dimensions.

As the extra dimensions are compact, one expects on general grounds that
the effects of the extra dimensions will be exponentially suppressed in
$\lowenspace$ at scales much longer than all the length scales of the
extra dimensions. While there exist parameter ranges where the
circumference of the extra dimension on the brane is arbitrarily large
compared with the orthogonal distance from the brane to the bolt,
our numerical
experiments have not uncovered parameter ranges where the distance to
the bolt could be made arbitrarily large compared with the circumference
on the brane. Our spacetimes are thus likely to produce an
observationally interesting correction to Newton's law only at length
scales that are comparable to the circumference of the extra dimension
on the brane.

The closing of the bulk at the bolts has two main appealing
consequences. First, as the extra dimensions are compact, the bulk has
no horizons that could develop singularities upon addition of
gravitational waves or black holes. For example, the
nonlinear gravitational wave of section \ref{sec:nonlinwave}
is manifestly regular in the bulk. Second,
the bolts make the brane world possible
without negative tension branes,
whereas with a periodic transverse dimension there would need to
be at least one negative tension
brane~\cite{leblond-myers-winters,gibbons-kallosh-linde}.
One could put our brane at an orbifold by identifying
the two sides of the bulk, provided such an orbifold is considered an
acceptable classical solution, but there seems to be a motivation to do so
only in the cases where the brane has negative tension.
It might be interesting to
investigate whether our spacetimes with a negative tension brane on an
orbifold
exhibit nonlinear instabilities
similar to those found in~\cite{marolf-trodden}.

As $\lowenspace$ may be any Einstein space, our construction includes
the Schwarzschild black hole on the brane,
as well as
dynamical vacuum black hole solutions on the brane. Such solutions offer
an arena for examining how the extra dimensions affect nonlinear
gravitational effects and black hole thermodynamics.
For example, one could ask whether bulk geodesics around black holes
can form halos near
the brane~\cite{chamblin-capture}.

Low energy phenomenological matter could be added
by a Lagrangian that is confined to the
brane~\cite{RS2}. The bulk Maxwell field need a priori not couple to any
such low energy matter. However, one might wish to investigate whether
coupling the bulk Maxwell field to charges on the brane induces an
effective $(p+1)$--dimensional Maxwell theory on~$\lowenspace$.

The dilatonic thick brane solutions of Gibbons and Maeda \cite{GM}
suggest possible generalizations of our brane worlds to include a bulk
dilaton field. As supergravity theories give rise to dilatonic bosonic
sectors, this raises the possibility of recovering bolt-brane-bolt
solutions to $M$-theory or to one of its supergravity limits. One may
hope to analyse such solutions in terms of $M$-theoretic (A)dS/CFT
correspondence \cite{malda-ads,witten-ads,witten-ds,strominger-ds}.
Since the extra dimensions form a two-dimensional submanifold the
problem of realising the solutions in a supergravity context may prove
to be simpler than the analogous problem in the Randall--Sundrum model.
\bigskip
\acknowledgments

We thank Gary Gibbons for valuable discussions, and in
particular for suggesting the non--linear wave construction of section
\ref{sec:nonlinwave}
to us.
We also thank John Barrett, Carsten Gundlach,
Kirill Krasnov, Eric Poisson and Jim
Vickers for
helpful comments.
For financial support,
DLW acknowledges Australian Research Council
Grant F6960043 and both authors acknowledge
Adelaide University Small Grant 21060100.
For hospitality during this work,
JL thanks the Department of Physics and Astronomy at
the University of Canterbury,
and DLW thanks the School of Mathematical Sciences at the University of
Nottingham and the Centre for Theoretical Physics at the University of
Sussex.


\begin{thebibliography}{999}
\bibitem{Ak}
K.~Akama,
{\it``Pregeometry''}, in K.~Kikkawa, N.~Nakanishi and H.~Nariai (eds),
{\it``Gauge Theory and Gravitation''},
{\it Lect.\ Notes Phys.\ } {\bf 176} (1982) 267 [\hepth{0001113}].

\bibitem{RuSha} V.~A. Rubakov and M.~E. Shaposhnikov,
{\it``Do we live inside a domain wall?''},
\plb{125}{1983}{136}.

\bibitem{Vis} M.~Visser,
{\it``An exotic class of Kaluza-Klein models''},
\plb{159}{1985}{22} [\hepth{9910093}];
\\
E.~J. Squires,
{\it``Dimensional reduction caused by a cosmological constant''},
\plb{167}{1986}{286}.

\bibitem{GW} G.~W. Gibbons and D.~L. Wiltshire,
{\it``Spacetime as a membrane in higher dimensions''},
\npb{287}{1987}{717} [\hepth{0109093}].

\bibitem{RS2}
L.~Randall and R.~Sundrum,
{\it``An alternative to compactification''},
\prl{83}{1999}{4690} [\hepth{9906064}].

\bibitem{GT}
J.~Garriga and T.~Tanaka,
{\it``Gravity in the brane-world''},
\prl{84}{2000}{2778} [\hepth{9911055}].

\bibitem{GKR}
S.~B. Giddings, E.~Katz and L.~Randall,
{\it``Linearized gravity in brane backgrounds''},
\jhep{0003}{2000}{023} [\hepth{0002091}].

\bibitem{ZK}
Z.~Kakushadze,
{\it``Gravity in Randall-Sundrum brane-world revisited''},
\plb{497}{2001}{125} [\hepth{0008128}].

\bibitem{CHR}
A. Chamblin, S.W. Hawking and H.S. Reall,
{\it``Brane-world black holes''},
\prd{61}{2000}{065007} [\hepth{9909205}].

\bibitem{emp-horo-myers}
R.~Emparan, G.~T. Horowitz and R.~C. Myers,
{\it``Exact description of black holes on branes''},
\jhep{0001}{2000}{007} [\hepth{9911043}].

\bibitem{emp-horo-myers2}
R.~Emparan, G.~T. Horowitz and R.~C. Myers,
{\it ``Exact description of black holes on branes. 2. Comparison with
BTZ black holes and black strings''},
\jhep{0001}{2000}{021} [\hepth{9912135}].

\bibitem{emp-horo-myers3}
R.~Emparan, G.~T. Horowitz and R.~C. Myers,
{\it ``Black holes radiate mainly on the brane''},
\prl{85}{2000}{499} [\hepth{0003118}].

\bibitem{chamblin-reall-etal}
A.~Chamblin, H.~S. Reall H.~Shinkai, and T.~Shiromizu,
{\it ``Charged brane world black holes''},
\prd{63}{2001}{064015} [\hepth{0008177}].

\bibitem{chamblin-capture}
A.~Chamblin,
{\it``Capture of bulk geodesics by brane world black holes''},
\cqg{18}{2001}{L17} [\hepth{0011128}]

\bibitem{emp-gregory-santos}
R.~Emparan, R.~Gregory and C.~Santos,
{\it``Black holes on thick branes''},
\prd{63}{2001}{104022} [\hepth{0012100}].

\bibitem{bridgman-malik-wands}
H.~A. Bridgman, K.~A. Malik and D.~Wands,
{\it``Cosmological perturbations in the bulk and on the brane''},
\prd{65}{2002}{043502} [\astroph{0107245}].

\bibitem{chodos-poppitz}
A.~Chodos and E.~Poppitz,
{\it ``Warp factors and extended sources in two transverse dimensions''},
\plb{471}{1999}{119} [\hepth{9909199}].

\bibitem{gregory}
R.~Gregory,
{\it ``Nonsingular global string compactifications''},
\prl{84}{2000}{2564} [\hepth{9911015}].

\bibitem{chacko-nelson}
Z.~Chacko and A.~E. Nelson,
{\it ``A solution to the hierarchy problem with an infinitely large
extra dimension and moduli stabilization''},
\prd{62}{2000}{085006} [\hepth{9912186}].

\bibitem{chen-luty-ponton}
J.-W. Chen, M.~A. Luty and E.~Ponton,
{\it ``A critical cosmological constant from millimeter extra
dimensions''},
\jhep{0009}{2000}{012} [\hepth{0003067}].

\bibitem{ola-vile}
I.~Olasagasti and A.~Vilenkin,
{\it ``Gravity of higher-dimensional global defects''},
\prd{62}{2000}{044014} [\hepth{0003300}];
\\
I.~Olasagasti,
{\it ``Gravitating global defects:
the gravitational field and compactification''},
\prd{63}{2001}{124016} [\hepth{0101203}].

\bibitem{gher-shapo}
T.~Gherghetta and M.~Shaposhnikov,
{\it ``Localizing gravity on a string-like defect in six dimensions''},
\prl{85}{2000}{240} [\hepth{0004014}].

\bibitem{gher-roe-shapo}
T.~Gherghetta, E.~Roessl and M.~Shaposhnikov,
{\it ``Living inside a hedgehog: higher dimensional solutions that
localize gravity''},
\plb{491}{2000}{353} [\hepth{0006251}].

\bibitem{haya-iza1}
S.~Hayakawa and K.~I. Izawa,
{\it ``Warped compactification with an abelian gauge theory''},
\plb{493}{2000}{380} [\hepth{0008111}].

\bibitem{ponton-poppitz}
E.~Ponton and E.~Poppitz,
{\it ``Gravity localization on string-like
defects in codimension two and the AdS/CFT correspondence''},
\jhep{0102}{2001}{042} [\hepth{0012033}].

\bibitem{corra-kaku}
O.~Corradini and Z.~Kakushadze,
{\it ``A solitonic three-brane in 6-D bulk''},
\plb{506}{2001}{167} [\hepth{0103031}].

\bibitem{kanti-madden-olive}
P.~Kanti, R.~Madden and K.~A. Olive,
{\it ``A 6-D brane world model''},
\prd{64}{2001}{044021} [\hepth{0104177}].

\bibitem{haya-iza2}
S.~Hayakawa and K.~I. Izawa,
{\it ``Warped compactification with a four-brane''},
Prog.\ Theor.\ Phys.\ {\bf106} (2001) 641 [\hepth{0106101}].

\bibitem{leblond-myers-winters}
F.~Leblond, R.~C. Myers and D.~J. Winters,
{\it ``Consistency conditions for brane
worlds in arbitrary dimensions''},
\jhep{0107}{2001}{031} [\hepth{0106140}];
\\
{\it ``Brane world sum rules and the AdS soliton''},
\hepth{0107034}.

\bibitem{kogan-etal}
I.~I. Kogan, S.~Mouslopoulos, A.~Papazoglou and G.~G. Ross,
{\it ``Multigravity in six dimensions: generating bounces with flat
positive tension branes''},
\prd{64}{2001}{124014} [\hepth{0107086}].

\bibitem{corradini-etal}
O.~Corradini, A.~Iglesias, Z.~Kakushadze and P.~Langfelder,
{\it ``Gravity on a 3-brane in 6D Bulk''},
\plb{521}{2001}{96} [\hepth{0108055}].

\bibitem{gubser-ads/cft}
S.~S. Gubser,
{\it``AdS/CFT and gravity''},
\prd{63}{2001}{084017} [\hepth{9912001}].

\bibitem{odintsov-cft}
S.~Nojiri, S.~D. Odintsov and S.~Zerbini,
{\it``Quantum (in)stability of dilatonic
AdS backgrounds and holographic renormalization group with gravity''},
\prd{62}{2000}{064006} [\hepth{0001192}];
\\
S.~Nojiri and S.~D. Odintsov,
{\it``Brane world inflation induced by quantum effects''},
\plb{484}{2000}{119} [\hepth{0004097}];
\\
S.~Nojiri, O.~Obregon and S.~D. Odintsov,
{\it``(Non)-singular brane-world
cosmology induced by quantum effects in d5 dilatonic gravity''},
\prd{62}{2000}104003{} [\hepth{0005127}].

\bibitem{Haw-Her-Reall}
S.~W. Hawking, T.~Hertog and H.~S. Reall,
{\it``Brane New World''},
\prd{62}{2000}{043501} [\hepth{0003052}].

\bibitem{BCR}
D.~Brecher, A.~Chamblin and H.~S. Reall,
{\it``AdS/CFT in the infinite momentum frame''},
\npb{607}{2001}{155} [\hepth{0012076}].

\bibitem{CG}
A.~Chamblin and G.~W. Gibbons,
{\it``Nonlinear supergravity on a brane without compactification''},
\prl{84}{2000}{1090} [\hepth{9909130}].

\bibitem{GH}
G.~W. Gibbons and S.~W. Hawking,
{\it``Classification of gravitational instanton symmetries''},
\cmp{66}{1979}{291}.

\bibitem{wetterich}
C.~Wetterich,
{\it``The cosmological constant and noncompact internal spaces in
Kaluza-Klein theories''},
\npb{255}{1985}{480}.

\bibitem{Mel}
M.~A. Melvin,
{\it``Pure magnetic and electric geons''},
{\it Phys. Lett.} {\bf8} (1963) 65;
\\
{\it``Dynamics of cylindrical electromagnetic universes''},
\pr{139}{1965}{B225}.

\bibitem{Kip} K.~S. Thorne,
{\it``Energy of infinitely long, cylindrically symmetric systems in general
relativity''},
\pr{138}{1965}{B251};
\\
{\it``Absolute stability of Melvin's magnetic universe''},
\pr{139}{1965}{B244}.

\bibitem{GM}
G.~W. Gibbons and K.~Maeda,
{\it``Black holes and membranes in higher
dimensional theories with dilaton
fields''},
\npb{298}{1988}{741}.

\bibitem{saffin}
P.~M. Saffin,
{\it ``Gravitating fluxbranes''},
\prd{64}{2001}{024014} [\grqc{0104014}].

\bibitem{GS}
M. Gutperle and A. Strominger,
{\it ``Fluxbranes in string theory''},
\jhep{0106}{2001}{035} [\hepth{0104136}].

\bibitem{CHC}
M.~S. Costa, C.~A.~R. Herdeiro and L. Cornalba,
{\it ``Fluxbranes and the dielectric effect in string theory''},
\npb{619}{2001}{155} [\hepth{0105023}].

\bibitem{uranga}
A.~M. Uranga,
{\it ``Wrapped fluxbranes''},
\hepth{0108196}.

\bibitem{fermions}
B.~Bajc and G.~Gabadadze,
{\it``Localization of matter and cosmological constant on a brane in
\adS\ space''},
\plb{474}{2000}{282} [\hepth{9912232}];
\\
I.~Oda,
{\it``Localization of matters on a string--like defect''},
\plb{496}{2000}{113}
[\hepth{0006203}];
\\
S.~Randjbar-Daemi and M.~Shaposhnikov,
{\it``Fermion zero-modes on brane-worlds''},
\plb{492}{2000}{361} [\hepth{0008079}].

\bibitem{israel-shell}
W.~Israel,
{\it ``Singular hypersurfaces and thin shells in general
relativity''},
\nc{B44}{1966}{1} [erratum: {\it ibid.}\ {\bf B48} (1966) 463].

\bibitem{MTW}
C.~W. Misner,
K.~S. Thorne and
J.~A. Wheeler,
{\it Gravitation\/}
(Freeman, San Francisco, 1973).

\bibitem{barrabes-israel}
C.~Barrabes and W.~Israel,
{\it ``Thin shells in general relativity and cosmology:
The lightlike limit''},
\prd{43}{1991}{1129}.

\bibitem{mukohyama-pertgen}
S.~Mukohyama,
{\it ``Perturbation of junction condition and doubly gauge-invariant
variables''},
\cqg{17}{2000}{4777} [\hepth{0006146}].

\bibitem{GV}
D.~Garfinkle and T.~Vachaspati,
{\it``Cosmic string traveling waves''},
\prd{42}{1990}{1960}.

\bibitem{BKL}
V.~Balasubramanian, P.~Kraus and A.~Lawrence,
{\it ``Bulk vs.\ boundary dynamics in anti-de Sitter spacetime''},
\prd{59}{1999}{046003} [\hepth{9805171}].

\bibitem{podolsky}
J.~Podolsky,
{\it``Interpretation of the
Siklos solutions as exact gravitational waves
in the anti-de Sitter universe''},
\cqg{15}{1998}{719} [\grqc{9801052}].

\bibitem{abra-stegun}
M.~Abramowitz
and
I.~A. Stegun (editors),
{\it Handbook of Mathematical Functions\/}
(Dover, New York, 1965).

\bibitem{dunfordII}
N.~Dunford and J.~S. Schwartz,
{\it Linear Operators\/}
(Interscience, New York, 1963),
Vol.~II.

\bibitem{reed-simonII}
M.~Reed and B.~Simon,
{\it Methods of Modern Mathematical Physics\/}
(Academic, New York, 1975),
Vol.~II.

\bibitem{gibbons-kallosh-linde}
G.~Gibbons, R.~Kallosh and A.~Linde,
{\it ``Brane world sum rules''},
\jhep{0101}{2001}{022} [\hepth{0011225}].

\bibitem{marolf-trodden}
D.~Marolf and M.~Trodden,
{\it ``Black Holes and Instabilities of Negative Tension Branes''},
\prd{64}{2001}{065019} [\hepth{0102135}].

\bibitem{malda-ads}
J.~Maldacena,
{\it ``The large N limit of superconformal field theories and supergravity''},
\atmp{2}{1998}{231} [\hepth{9711200}].

\bibitem{witten-ads}
E.~Witten,
{\it ``Anti-de~Sitter space and holography''},
\atmp{2}{1998}{253} [\hepth{9802150}].

\bibitem{witten-ds}
E.~Witten,
{\it ``Quantum gravity in de~Sitter space''},
\hepth{0106109}.

\bibitem{strominger-ds}
A.~Strominger,
{\it ``The dS/CFT correspondence''},
\jhep{0110}{2001}{034} [\hepth{0106113}].

\end{thebibliography}
\end{document}